\documentclass[prx, twocolumn, superscriptaddress,
 showpacs,
 longbibliography,
 aps,
 showkeys]{revtex4-2}
\usepackage[utf8]{inputenc}
\usepackage{graphicx} 
\usepackage{physics}
\usepackage{float}
\usepackage[colorlinks=true,allcolors=blue]{hyperref}

\def\equationautorefname#1#2\null{Eq.#1(#2\null)}

\begin{document}
\title{Photon Antibunching in Single-Molecule Vibrational Sum-Frequency Generation}
\author{Fatemeh Moradi Kalarde}
\affiliation{Institute of Physics, Swiss Federal Institute of Technology Lausanne (EPFL), CH-1015 Lausanne, Switzerland} 
\affiliation{Inria Paris-Saclay and CPHT, Ecole polytechnique, Institut Polytechnique de Paris, France}
\author{Carlos Sánchez Muñoz}
\affiliation{Condensed Matter Physics Center (IFIMAC), Universidad Autónoma de Madrid, Spain}
\affiliation{Institute of Fundamental Physics IFF-CSIC, Calle Serrano 113b, 28006 Madrid, Spain}
\author{Johannes Feist}
\affiliation{Condensed Matter Physics Center (IFIMAC), Universidad Autónoma de Madrid, Spain}
\affiliation{Departamento de Física Teórica de la Materia Condensada, Universidad Autónoma de Madrid, Spain}
\author{Christophe Galland}
\affiliation{Institute of Physics, Swiss Federal Institute of Technology Lausanne (EPFL), CH-1015 Lausanne, Switzerland} 
\affiliation{Center of Quantum Science and Engineering, Swiss Federal Institute of Technology Lausanne (EPFL), CH-1015 Lausanne, Switzerland
}%
\date{\today}

\begin{abstract}
Sum-frequency generation (SFG) allows for coherent upconversion of an electromagnetic signal and has applications in mid-infrared vibrational spectroscopy of molecules. Recent experimental and theoretical studies have shown that plasmonic nanocavities, with their deep sub-wavelength mode volumes, may allow to obtain vibrational SFG signals from a single molecule. 
In this article, we compute the degree of second order coherence ($g^{(2)}(0)$) of the upconverted mid-infrared field under realistic parameters and accounting for the anharmonic potential that characterizes vibrational modes of individual molecules.
On the one hand, we delineate the regime in which the device should operate in order to preserve the second-order coherence of the mid-infrared source, as required in quantum applications. On the other hand, we show that an anharmonic molecular potential can lead to antibunching of the upconverted photons under coherent, Poisson-distributed mid-infrared and visible drives. Our results therefore open a path toward a new kind of bright and tunable source of indistinguishable single photons by leveraging ``vibrational blockade'' in a resonantly and parametrically driven molecule, without the need for strong light-matter coupling.


\end{abstract}

\maketitle

\section{Introduction}

Single-photon sources are a key resource for quantum technologies~\cite{eisaman2011invited} with pivotal applications in quantum computation~\cite{knill2001}, communications~\cite{su2009}, and metrology~\cite{muller2017}. The main challenge in this area is to realize a high-rate solid-state photon source producing indistinguishable photons~\cite{santori2002indistinguishable}, which demands that pure-dephasing makes a negligible contribution to the emission linewidth~\cite{kuhlmann2015transform}. Probabilistic approaches to single-photon emission typically leverage an optical nonlinearity in the material (spontaneous parametric down conversion~\cite{zhang2021spontaneous} or degenerate four-wave mixing~\cite{garay2023fiber}) to generate time-correlated photon pairs from bulk crystals~\cite{burnham1970spdc} or photonic integrated circuits~\cite{wang2021integrated}. The detection of one photon in, e.g., the idler mode heralds a quantum state very close to a single-photon Fock state in the signal mode~\cite{eisaman2011invited}. This heralding works provided that the probability of photon pair generation per mode is well below unity, a condition that fundamentally limits the brightness of these type of sources
~\cite{esmann2024solid}.
An alternative approach exploits the anharmonic character of quantum emitters to achieve deterministic single-photon emission via photon blockade~\cite{tian1992quantum,birnbaum2005photon}. Such emitters include trapped atoms or ions~\cite{kimble1977photon}, and notably solid-state sources such as immobilized molecules~\cite{basche1992photon}, quantum dots~\cite{arakawa2020reviewQD}, nanotubes~\cite{hogele2008photon}, color centers~\cite{brouri2000photon,kurtsiefer2000stable}, and many other rising low-dimensional materials~\cite{esmann2024solid}.

\begin{figure*}
    \centering
    \includegraphics[width=0.95\textwidth]{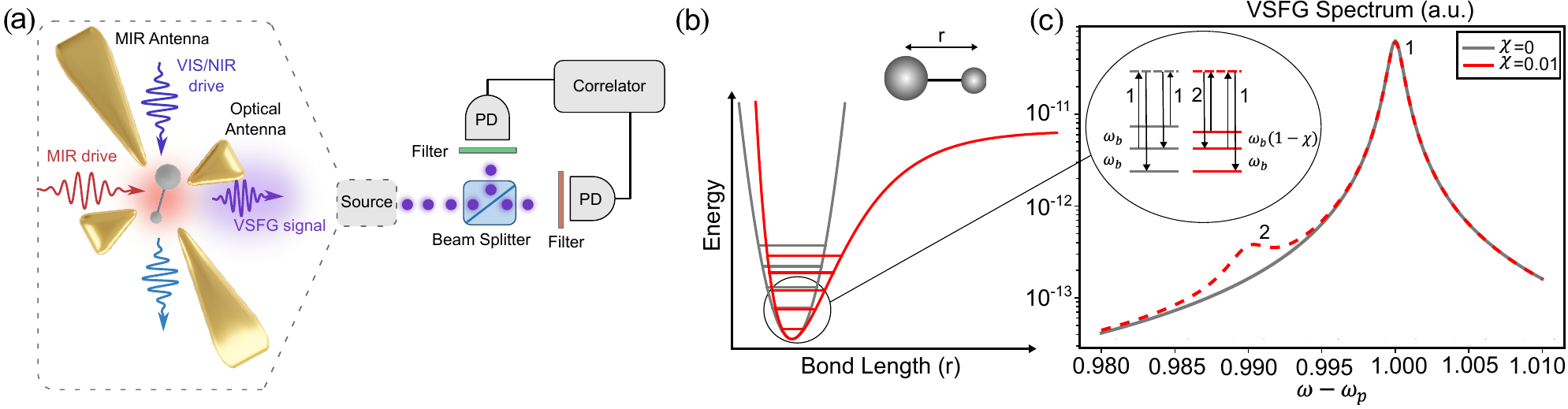}
    \caption{\textbf{Overview of the proposed scheme.} (a) VIS and MIR antennas mediate an efficient interaction between photonic and molecular vibrational modes. A spectrometer or HBT interferometer measures the spectrum or the frequency-filtered second-order correlation of the scattered photons respectively. (b) Vibrational potential of the molecule considering either a Morse potential (red line) or its harmonic approximation (grey line). (c) As a result of anharmonicity, a second peak emerges in the thermally activated, spontaneous anti-Stokes Raman spectrum.}
    \label{fig:intro}
\end{figure*}

In the last decades, molecules have emerged as a particularly promising platform for integrated quantum photonic technologies~\cite{toninelli2021}, given their good degree of coherence~\cite{zadrozny2015}, frequency tunability~\cite{colautti2020,lange2024}, versatile coupling to photonic structures~\cite{faez2014,chikkaraddy2016,wang2017a,grandi2019}
and strong non-linearity at the single-photon level~\cite{wang2019} enabling effects such as the emission of antibunched light~\cite{wrigge2008,lombardi2018}, four-wave mixing~\cite{pscherer2021} and sum-frequency generation~\cite{roelli2020molecular,chen2021continuous}.

Here, we propose a new implementation of a bright source of single photons based on photon blockade in vibrational sum-frequency generation (VSFG) from a single molecule embedded in a dual-resonant plasmonic nanocavity in the weak-coupling regime~\cite{roelli2020molecular,chen2021continuous}, see Fig.~\ref{fig:intro}(a). In contrast to a recent proposal for unconventional photon blockade in a hybrid plasmonic-photonic molecular optomechanical cavity~\cite{abutalebi2024}, our scheme is operational for arbitrarily weak molecule-cavity coupling strengths and irrespective of the cavity modes quality factors. 
VSFG is a coherent upconversion process that occurs whenever a vibrational mode that is both Raman- and infrared-active (which requires a non-centrosymmetric molecule) is driven simultaneously with a mid-infrared (MIR) laser tuned on resonance with the vibrational frequency and a visible (VIS) or near-infrared (NIR) laser, typically tuned away from any electronic resonance to avoid parasitic incoherent fluorescence~\cite{shen1989surface,wang2005quantitative,lambert2005implementing}. Three-wave mixing results in upconverted photons at frequencies corresponding to the sum and difference of the incoming laser frequencies; while the difference frequency signal is polluted by spontaneous Raman scattering, VSFG is almost free of incoherent background due to the very low thermal occupancy of the vibration at MIR frequencies, even at room temperature. We will show the feasibility of single-photon generation using VSFG from a single molecule with a sufficient anharmonicity in one of its vibrational modes, resulting in antibunching in the sum-frequency upconverted photons. 

Our proposal shares with resonance fluorescence some nuances regarding the coexistence of coherence (sub-natural linewidth) and single-photon emission recently explored in several works~\cite{carreno2018,zubizarreta2020conventional,phillips2020,hanschke2020}. Antibunching is revealed at specific frequency windows through the zero-delay second-order correlation function of the frequency-filtered emission~\cite{delvalle_2012,gonzalez-tudela2013a,schmidt_esteban_gonzaleztudela_giedke_aizpurua_2016}, which, when analyzed over the parameter phase  space, unveils a complex landscape of interference phenomena between the coherent VSFG nonlinear process and the incoherent anti-Stokes background. This interference is crucial for enabling single-photon emission, which is maximized when the MIR laser is resonant with the vibrational mode and the coherent and incoherent peaks overlap. These results establish our proposed molecular scheme as a novel, privileged platform for studying the role of interference in single-photon emission, closely tied to the notions of conventional and unconventional photon blockade~\cite{zubizarreta2020conventional}. 

The versatility of our scheme, together with the picosecond relaxation times of vibrational excitations, promises a valuable tool for the production of indistinguishable single photons at THz repetition rates, enabling a breadth of applications such as the realisation of multi-photon entanglement in boson sampling~\cite{brod2019photonic} and cluster states~\cite{istrati2020sequential,tiurev2022high,cogan2023deterministic,ferreira2024deterministic}. Furthermore, our observation of the rich phenomenology imprinted in the photon correlations by the nonlinear molecular dynamics suggests promising prospects for quantum Raman spectroscopy boosting analytical, material, and biomedical applications~\cite{orlando2021a,kudelski2008a,qian2008a,yang2011a}; quantum estimation of molecular parameters and quantum states~\cite{gambetta2001}; and quantum sensing~\cite{yu2021}.

In the following, we provide a numerical analysis of VSFG based on the scenario presented in~\cite{chen2021continuous,
roelli2020molecular} in the single or few-molecule limit, focusing on a more rigorous treatment of classical and quantum coherence in the upconversion process and introducing the anharmonicity of the vibrational mode~\cite{duncan1991determination}. For the interaction of the molecular vibration with visible light we adopt an optomechanical description of Raman scattering~\cite{roelli2016molecular, schmidt2017linking, esteban_baumberg_aizpurua_2022}. We first confirm that in the harmonic approximation for the vibrational potential the first- and second-order coherence of the infrared signal are preserved in the upconverted signal, as required for quantum frequency conversion~\cite{zaske2012visible,rutz2017quantum} and MIR single-photon spectroscopy~\cite{temporao2006mid,dam2012room,tang2015two,barh2017ultra,barh2018thermal,lehmann2019single,huang2021mid,huang2022wide,wang2023mid,cai2024mid,li2024ultra}. Second, we demonstrate the feasibility of generating antibunched photons under coherent MIR and VIS excitation of a single molecule having a strong enough anharmonicity in its vibrational potential~\cite{vento2023measurement}, while the molecule remains in the weak coupling regime with both MIR and VIS cavity modes. We conclude by exploring the effect of the number of involved molecules on the degree of antibunching.

\section{Results}
\paragraph{Model}
We consider a single vibrational mode of a single molecule placed in a dual-resonant plasmonic cavity featuring two resonances: one in the MIR at $\omega_\text{IR}\simeq \omega_{\rm b}$, where $\omega_{\rm b}$ is the transition frequency between the ground and first vibrational excited state, and one in the VIS/NIR domain at frequency $\omega_{\rm c}$ that is considered much higher than the vibrational frequency and sufficiently lower than the first electronic transition of the molecule. 
The nanocavity is driven by two continuous monochromatic laser beams: one at a MIR frequency $\omega_\text{d}$ near-resonant with the cavity mode and molecular vibration, $\omega_\text{d}\simeq\omega_\text{IR}\simeq\omega_{\rm b}$, and one at a VIS/NIR frequency $\omega_{\rm p}\simeq\omega_{\rm c}$, generating a VSFG signal at a frequency $\omega_\mathrm{VSFG}\equiv \omega_\text p + \omega_\text d$. The setup is depicted in \autoref{fig:intro}a.

The Hamiltonian of the molecule, restricted to the ground electronic state and the few first vibrational states, is expressed as:
\begin{align}
\label{eq:MoleculeHamilton}
 H_{\rm mol} &= \omega_{\rm b} b^{\dagger} b - \frac{\chi}{2} \omega_b b^{\dagger}b^{\dagger} b b,
\end{align}
Here and in the following, we choose units where the reduced Planck constant $\hbar=1$. The vibrational (phonon) of the molecule at frequency  $\omega_{\rm b}$ has a corresponding annihilation operator denoted as $b$. The parameter $\chi$ characterizes the anharmonicity of the vibrational mode. The eigenstates of $H_{\rm mol}$ are the Fock states $b^{\dagger} b \ket{n} = n \ket{n}$, with energies given by $E_n = n \omega_{\rm b} \left[1 - \chi  (n-1)/2 \right]$.  We will restrict ourselves to small, realistic values $\chi\leq 10^{-2}$~\cite{vento2023measurement}, for which the anharmonicity is only a small correction to the eigenenergies of the first few excited states. A recent theoretical extension of molecular optomechanics discussed more comprehensively how to treat the vibrational anharmonicity \cite{schmidt2024}. 

A simple Hamiltonian can be formulated to describe the full system, which takes into account the molecular vibration, the two resonances of the nanocavity, the incident lasers ($\Omega_{\rm IR}$ and $\Omega_{\rm c}$ denote their respective drive strengths), as well as the resonant dipolar coupling between the MIR cavity mode and the molecular vibration, and its optomechanical (Raman) coupling with the VIS/NIR mode~\cite{roelli2020molecular}
\begin{equation}
\label{eq:fullHamilton}
\begin{split}
H &= H_{\rm mol} + \omega_{\rm IR} a_{\rm IR}^{\dagger} a_{\rm IR} + \omega_{\rm c} a_{\rm c}^{\dagger} a_{\rm c}\\
  &+ i \Omega_{\rm IR}(a_{\rm IR}^{\dagger} e^{-i \omega_{\rm d} t}-a_{\rm IR}e^{i \omega_{\rm d} t}) \\
  &+ i \Omega_{\rm c}(a_{\rm c}^{\dagger} e^{-i \omega_{\rm p} t}-a_{\rm c}e^{i \omega_{\rm p} t})\\
  &+ \tilde{g}_{\rm IR} (a_{\rm IR}^{\dagger} +a_{\rm IR} )(b^{\dagger} + b) + \tilde{g}_{\rm c} a_{\rm c}^{\dagger} a_{\rm c} (b^{\dagger} + b) 
\end{split}
\end{equation}

We assume decay rates of $\kappa_{\rm c}$, $\kappa_{\rm IR}$, and $\gamma$ for the optical, infrared, and phonon modes, respectively.
Following the approximations detailed in the Appendix, we obtain the linearized Hamiltonian
\begin{equation}
\label{eq:linearizedHamilton}
\begin{split}
H &= H_{\rm mol} + (\omega_{\rm c}- \omega_{\rm p})\delta a_{\rm c}^{\dagger} \delta a_{\rm c}\\
 &+ \tilde{g}_{\rm c} ( \alpha_{\rm c} \delta a_{\rm c}^{\dagger} + \alpha_{\rm c}^{\star} \delta a_{\rm c} )(b^{\dagger} + b )\\
 &+ \tilde{g}_{\rm IR} (\alpha_{\rm IR}^{\star} b e^{i \omega_{\rm d} t} +\alpha_{\rm IR} b^{\dagger} e^{-i \omega_{\rm d} t})
\end{split}
\end{equation}
where $\alpha_{\rm IR} =\frac{\Omega_{\rm IR}}{i \Delta_{\rm IR} +\frac{ \kappa_{\rm IR} }{2}}$ and $ \alpha_{\rm c} =\frac{\Omega_{\rm c}}{i \Delta_{\rm c} + \frac{\kappa_{\rm c}}{2}}$ are the intracavity field amplitudes, with $\Delta_{\rm IR}=\omega_{\rm IR}-\omega_{\rm d}$ and $\Delta_{\rm c}=\omega_{\rm c}-\omega_{\rm p}$ the cavity-laser detunings.
The annihilation operator $\delta a_c$ denotes fluctuations in the VIS/NIR mode, while the MIR mode operator is eliminated and only the coherent amplitude $ \alpha_{\rm IR}$ is retained. The decay rates $\kappa_{\rm c}$ and $\gamma$ apply to $\delta a_c$ and $b$, respectively.

In case of zero detuning of the lasers from the cavity resonances (such that $\alpha_{\rm IR}$ and $\alpha_{c}$ are real), the Hamiltonian simplifies to
\begin{multline}
\label{eq: finalHamilton}
H = \omega_{\rm b} b^{\dagger} b - \frac{\chi}{2} \omega_\text b b^{\dagger} b^{\dagger} b b
  + g_{\rm c} ( \delta a_{\rm c}^{\dagger} + \delta a_{\rm c} )(b^{\dagger} + b )\\
  + g_{\rm IR} ( b e^{i \omega_{\rm d} t}+ b^{\dagger} e^{-i \omega_{\rm d} t}),
\end{multline}
where we have defined $g_{\rm IR} = \tilde{g}_{\rm IR} \alpha_{\rm IR}$ and $g_{c} = \tilde{g}_{c} \alpha_{c}$. 
We note that in the regime of vanishing VIS/NIR drive, this reduces to
\begin{equation}
\label{eq: IR_Regime_Hamiltonian}
H = \omega_{\rm b} b^{\dagger} b - \frac{\chi}{2} \omega_{\rm b} b^{\dagger} b^{\dagger} b b
   + g_{\rm IR} ( b e^{i \omega_{\rm d} t}+ b^{\dagger} e^{-i \omega_{\rm d} t}),
\end{equation}
showing that optical and vibrational modes are effectively decoupled.

The Lindblad master equation corresponding to the Hamiltonian in \autoref{eq: finalHamilton} with the Lindblad dissipation terms due to the decay rates mentioned above is solved using QuTiP~\cite{Johansson2012,Johansson2013}. 
The parameters are fixed in dimensionless units by normalizing all energies to the vibrational frequency $\omega_b$, which typically takes values of tens of THz localized molecular vibrations. 
The values set throughout the article are presented in \autoref{tab: param}, unless stated otherwise.  These values are chosen to match accessible experimental conditions~\cite{roelli2020molecular,chen2021continuous}. The dimensionless thermal occupancy is set as $n_{\rm th}=(e^{\frac{\omega_b}{k_{\mathrm{B}}T}}-1)^{-1}$, where $k_{\mathrm{B}}$ is the Boltzmann constant and $T$ is the temperature.
\begin{table}[b ]
\centering
\setlength{\tabcolsep}{2pt} 
\begin{tabular*}{\columnwidth}{@{\extracolsep{\fill}} c c c c c c c c c }  
\hline
 $\omega_b$ & $\omega_{\rm d}$ & $\gamma$  & $g_{\rm IR}$ & $n_{\rm th}$ & $\Delta_c$ & $g_{c}$ & $\kappa_c$   \\ \hline \hline
 1 & 1 & 0.001 & $10^{-5}$ & $10^{-4}$ & 0 & $10^{-5}$  & 4   \\ 
 \hline
\end{tabular*}
\caption{Default parameters used in the article, unless stated otherwise. } 
\label{tab: param}
\end{table}
Numerical resolution of the Lindblad equation requires to truncate the Hilbert sub-spaces of each bosonic mode; convergence was achieved for a large range of driving strengths with the following dimensions:
$q_c=4$ for the VIS/NIR cavity, $q_b=5$ for vibrational mode. Since the Hamiltonian is time-dependent but periodic, we obtain a Floquet expansion of the steady state, $\rho_s(t) = \sum_{n=-n_b}^{n_b} \rho_n e^{i n \omega_d t}$, truncated at $n_b=4$ terms.
When computing the spectrally resolved VSFG intensity, the linewidth of the two-level system acting as a filter is set to 
$\Gamma=10^{-5}$ and its coupling strength to $\epsilon= 10^{-6}$. 
These parameters are modified to 
$\Gamma=10^{-2}$ and $\epsilon= 10^{-5}$ when computing spectrally-resolved $g^{(2)}(0)$.


\begin{figure*}[t]
    \centering
    \includegraphics[width=\textwidth]{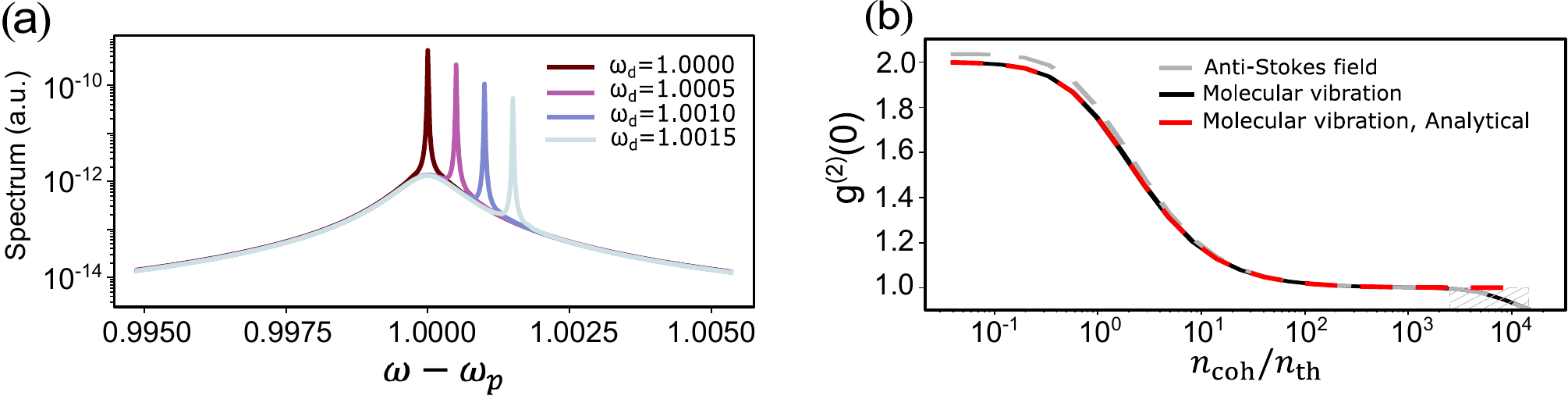}
    \caption{\textbf{VSFG under harmonic vibrational potential.} (a) Anti-Stokes spectra at various infrared laser frequencies. The sharp peak, with its frequency shifting in accordance with the MIR laser frequency, exhibits a linewidth that corresponds to the filter's characteristics.  (b) Second-order correlation function of vibrations and anti-Stokes photons plotted versus MIR drive strength, under the assumption of zero anharmonicity. Dashed area shows the interval where the truncated Hilbert space fails to approximate the infinite levels of harmonic oscillator correctly.}
    \label{fig:harmo}
\end{figure*}

\paragraph{Harmonic vibration}
We begin with results obtained under the harmonic potential approximation for the vibrational mode, i.e. $\chi=0$.
Vibrational Raman (anti-Stokes side of the VIS/NIR laser) and VSFG spectra are shown in \autoref{fig:harmo}a for various MIR laser frequencies. Each spectrum features a broad and MIR-laser-independent peak corresponding to thermally-activated spontaneous Raman scattering. For increasing VIS/NIR laser power, this peak can grow in intensity through vibrational pumping~\cite{kneipp1996,maher2008}, which is a consequence of quantum back-action in the formalism of molecular cavity optomechanics used here~\cite{benz2016}. In our study, we keep the VIS/NIR laser power low enough that this effect is negligible.

On top of the broad anti-Stokes Raman peak, the coherently upconverted VSFG signal at a frequency $\omega_\mathrm{VSFG} \equiv \omega_p + \omega_d$ can be seen. Since the driving lasers in our simulation have perfectly defined frequencies, the linewidth of this signal is set only by the linewidth of the detector (see Appendix), while its frequency is set by energy conservation to the sum of the MIR and VIS/NIR laser frequencies. This behavior was experimentally evidenced in Ref.~\citenum{chen2021continuous}. In practice, the lineshape of the VSFG signal is expected to be a convolution of those of the two driving lasers.

In order to quantify the photon statistics and second-order coherence properties of the upconverted field we compute the second-order correlation function at zero time delay, defined for a particular mode $j$ as:

\begin{equation}
\label{eq:2ndorder_Mode_b}
g^{(2)}_j(\tau=0) = \frac{\langle j^{\dagger}j^{\dagger}jj\rangle}{{\langle j^{\dagger}j\rangle}^2}
\end{equation}

In order to obtain frequency-resolved photon statistics of the emission of the system, we employ the method of Ref.~\citenum{delvalle_2012}, consisting of including explicit detectors with a Lorentzian frequency response in the simulation. This method is summarized in Appendix~\ref{subsec: freqFilteredMethod}.

\autoref{fig:harmo}b shows that as the MIR drive strength increases, both the vibrational mode and the VSFG field exhibit a transition from a thermal state to a coherent state. At low enough MIR drive, the vibrational mode is in thermal equilibrium and the anti-Stokes signal is dominated by spontaneous Raman scattering that inherits the same thermal statistics corresponding to $g^{(2)}=2$~\cite{tarragovelez2019}. As the MIR drive increases, the coherent contribution of VSFG to the displaced thermal state increases and $g^{(2)}$ asymptotically reaches 1. 


One can gain more insight into this behavior by considering the Hamiltonian of the vibrational mode in the rotating frame
\begin{equation}
\label{eq: displaced_HO}
H_{\rm b} =  (\omega_{\rm b}- \omega_{\rm d}) b^{\dagger} b
+ g_{\rm IR} ( b + b^{\dagger})
\end{equation}
The second term corresponds to a coherent displacement of the initial thermal state. 
In the special case of zero temperature ($n_{\rm th}=0$), the solution for the vibrational mode is a coherent state $\ket{\beta}$ of amplitude
\begin{equation}
\label{eq:average_b}
\beta= \frac{g_{\rm IR}}{- (\omega_{\rm b} - \omega_{\rm m})+i \frac{\gamma}{2} }.
\end{equation}
It is convenient to define a coherent population $n_{\rm coh}= \vert \beta \vert^2$, which corresponds to the average number of excitations in the vibrational mode when $g_c=0$ and $n_{\rm th}=0$.

In the general case of arbitrary $n_{\rm th}$, the mean vibrational population is given by
\begin{equation}
n_{\rm b}= \Tr{\rho_{\rm th}\, (b^{\dagger}+ \beta^{\star})(b+\beta)} = n_{\rm th} + {\vert \beta \vert}^{2},
\end{equation}
with
\begin{equation*}
\rho_{\rm th}= \sum_{n=0}^{\infty} P_n \ket{n}\bra{n} \quad \text{with} \quad P_n= \frac{n_{\rm th}^{n}}{(1+n_{\rm th})^{n+1}}.
\end{equation*}
The non-normalized second-order correlation is given by:
\begin{equation*}
\Tr{\rho_{\rm th}\, (b^{\dagger}+ \beta^{\star})^{2}(b+\beta)^{2} }
= 2 n_{\rm th}^{2} + 4 n_{\rm th}{\vert \beta \vert}^{2} + {\vert \beta \vert}^{4}
\end{equation*}
and after normalization
\begin{equation}
\label{eq: g2_zeroChi}
g^{(2)}_{\rm b}(0) = 2- \frac{{\vert \beta \vert} ^{4}}{(n_{\rm th} + {\vert \beta \vert}^{2})^{2}} 
= 2- \frac{1}{(1 + \frac{n_{\rm th}}{n_{\rm coh}})^{2}},
\end{equation}
which is the analytical expression plotted as a dashed red curve in \autoref{fig:harmo}b. The hashed area at high $n_{\rm coh}$ shows the domain where the Hilbert space truncation causes inaccuracies in the numerical solution. We therefore limit future calculations to MIR powers below this value.


\begin{figure}
    \centering
    \includegraphics[width=0.95\columnwidth]{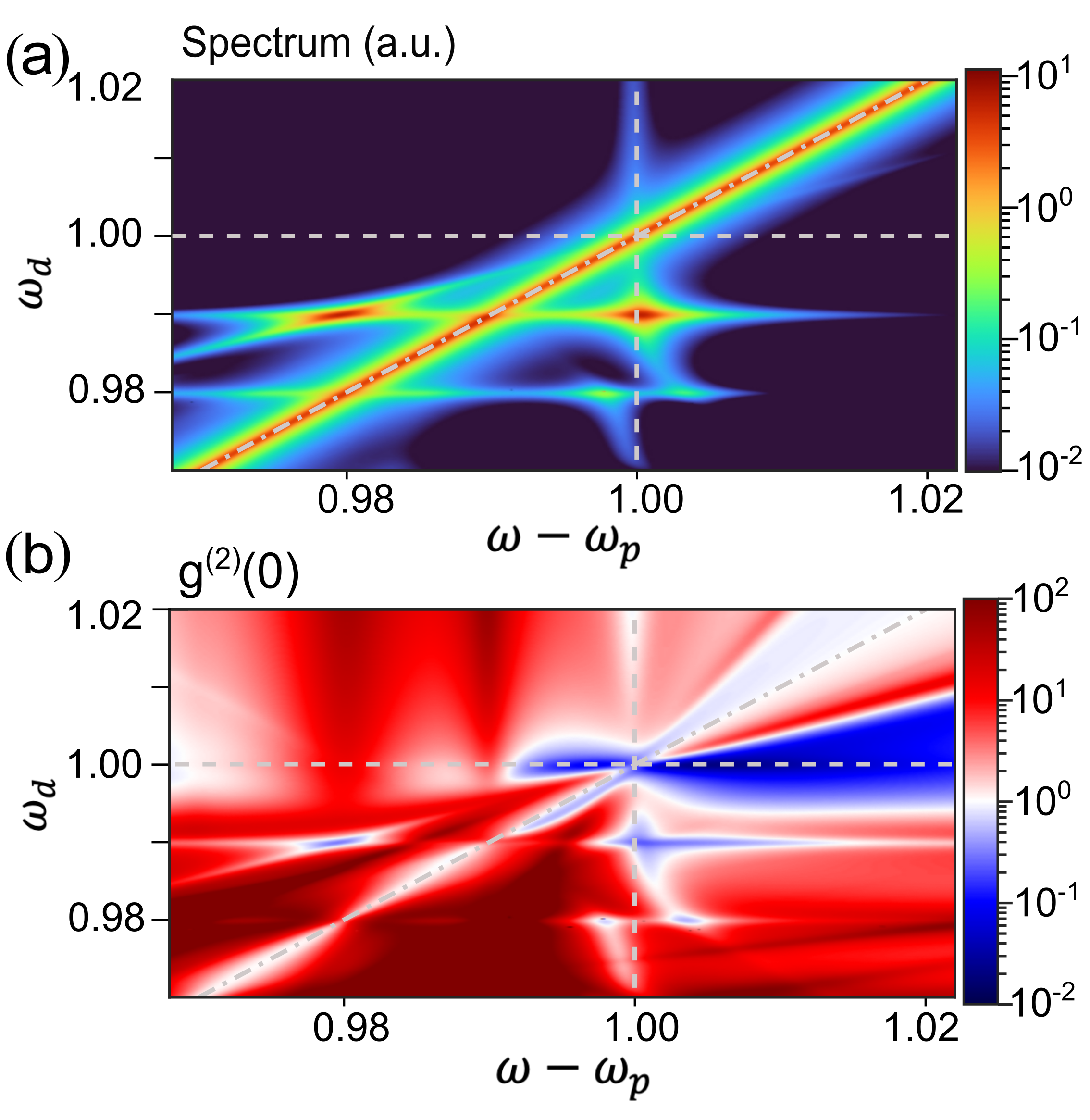}
    \caption{\textbf{VSFG under anharmonic vibrational potential.} Scan of the (a) emission intensity and (b) second-order correlation function as a function of driving frequency $\omega_d$ and emission frequency under strong MIR driving. The dot-dashed grey lines indicate $\omega_d = \omega - \omega_p$.}
    \label{fig:g2map}
\end{figure}

\paragraph{Anharmonicity in the vibration}

In reality, vibrational modes of small molecules have non-negligible anharmonicities~\cite{morichika2019molecular}, responsible for temperature-dependent Raman shifts and lineshapes~\cite{zhang2019identification} and for the observation of so called `hot bands' in Raman scattering when the anharmonicity is larger than the peak linewidth~\cite{vento2023measurement}.  

The main insight of this work is that the presence of this anharmonicity introduces rich features in the spectral properties of the emission, particularly in its statistics. This is shown in Fig.~\ref{fig:g2map}, where we plot both the spectrum and the frequency-resolved $g^{(2)}(\omega)$ as a function of the driving frequency $\omega_\text d$. 
Here, the anharmonicity parameter is set to $\chi=-0.02$ and filter linewidth is kept equal at $\Gamma=10^{-3}$ in both panels for direct comparison. The dot-dashed grey lines indicate the frequency of the VSFG signal, confirming that it remains the most important contribution to the emission. Nevertheless, the anharmonicity enables multiphoton excitation processes that imprint new features on the spectrum. In particular, when two-photon absorption to the second excited state is resonant, at $\omega_\text d = 0.99$, additional double-peaked emission appears at the transition frequencies between levels $1\to 0$ ($\omega-\omega_\text p=1$) and $2\to 1$ ($\omega-\omega_\text p=0.98$). 

Regarding photon statistics, the VSFG signal is observed to be antibunched provided that it overlaps with the incoherent anti-Stokes background, i.e., that the driving frequency is close to the vibrational transition, $\omega_\text{d} \approx \omega_\text b$. This establishes that the incoherent background due to the anti-Stokes process must be present to obtain strong antibunching, in accordance with the recent understanding of antibunching in resonance fluorescence as a consequence of interference between coherent and incoherent terms~\cite{carreno2018,zubizarreta2020conventional,phillips2020,hanschke2020}. A second prominent antibunching feature is observed under driving at the two-photon resonance $\omega_\text d=0.99$, which leads to antibunched emission at the two frequencies $\omega-\omega_p = \{0.98, 1\}$. Moreover, the frequency-resolved cross-correlation measured at these two frequencies reveals a large bunching, $g^{(2)}(\omega_1-\omega_p=1, \omega_2-\omega_p =0.98) \approx 8$, indicating that these photons are produced in pairs through cascaded emission. This could be exploited for the generation of entangled photon pairs in the VIS or NIR range, which could be potentially useful for quantum communication protocols.

\begin{figure}[t]
    \centering
    \includegraphics[width=0.95\columnwidth]{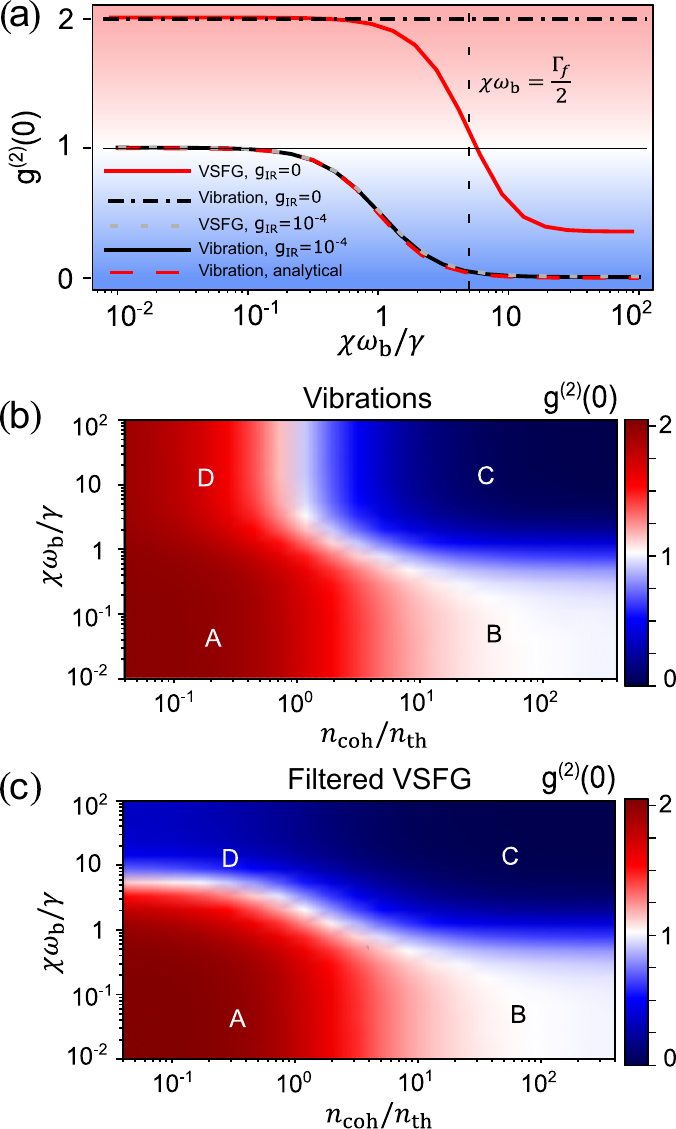}
    \caption{\textbf{Regimes of statistics of VSFG} (a) Evolution of $g^{(2)}$ of the vibrational mode and of the filtered anti-Stokes field around the sum-frequency plotted against the relative strength of anharmonicity, both in the absence ($g_\text{IR}=0$) and in the presence ($g_\text{IR}=10^{-4}$) of MIR drive. The analytical curve is from \autoref{eq: second order_b}. (b,c) Second-order correlation function of vibrations (b) and filtered anti-Stokes field (c) plotted versus the dimensionless strengths of MIR drive and anharmonicity. Capital letters identify four regions of parameters discussed in the text.}
    \label{fig:anhar}
\end{figure}

Further understanding of this phenomena and its dependence with the anharmonicity $\chi$ can be obtained by analyzing the fluctuations of the vibrational degree of freedom. Under the assumption of $\tilde{g}_{\rm IR} \ll \gamma$, $\tilde{g}_{\rm c} =0$ and $n_\text{th}=0$, the second-order coherence of the resonantly driven vibrational mode has an analytical expression (see, for example, Ref.~\citenum{zubizarreta2020conventional}):
\begin{equation}
\label{eq: second order_b}
    g^{(2)}_{\rm b}(0) = \frac{\gamma^{2} + 4(\omega_{\rm b}- \omega_{\rm m})^{2}}{\gamma^{2} + \left[2(\omega_{\rm b}- \omega_{\rm m})-\chi \omega_{\rm b}\right]^{2}}
\end{equation}
This expression is plotted in \autoref{fig:anhar}a against the dimensionless parameter $\chi \omega_{\rm b}/\gamma$, showing excellent agreement with the numerical computation of $g_\text{b}^{(2)}(0)$ using the full Lindblad equation. The figure also shows that the second-order correlation of the anti-Stokes field filtered around the VSFG signal under the antibunching condition $\omega_\text d = \omega_\text b$  very closely follows $g_\text{b}^{(2)}(0)$, as expected given the `beam-splitter' form of the optomechanical Hamiltonian for the anti-Stokes sideband (see Appendix~\ref{subsec: beamsplitter interaction}). Most importantly, all calculations confirm that when the anharmonicity is larger than the vibrational linewidth, $\chi \omega_{\rm b}\geq \gamma$, the filtered anti-Stokes signal becomes strongly antibunched. However, it should be noted that even though the main emission line is coherent and narrow (limited by the filter linewidth), antibunching is only observed if the emission from a sufficiently broad frequency interval is included (see \autoref{fig:g2_vs_chi_diff_Gamma_highIR}), i.e., if the filter is not too narrow and the incoherent background is included. Therefore, joint narrow linewidths and single-photon emission cannot be straightforwardly obtained, as has been recently illustrated in
resonance fluorescence in driven two-level systems~\cite{carreno2018,zubizarreta2020conventional,phillips2020,hanschke2020}.

This result should be compared with the curves obtained without MIR drive in \autoref{fig:anhar}a, corresponding to the case of spontaneous Raman scattering from a single molecule. In this case, while the vibrational mode remains in thermal equilibrium, it is also possible to observe photon antibunching in the spectrally filtered anti-Stokes field, but only under the intuitive condition that the anharmonicity exceeds the filter linewidth $\Gamma$. There are two other major drawbacks with this undriven approach: First, the spontaneous anti-Stokes photon flux can only be increased by increasing the VIS/NIR pump power, which more readily induces damage in the nanocavity~\cite{crampton2016}. Second, the first-order coherence property of the anti-Stokes field is dictated by that of the molecular vibration, for which the linewidth $\gamma$ can have non-negligible contributions from pure dephasing, thereby degrading the indistinguishability of the antibunched photons.

In order to capture the different regimes of photon statistics for the upconverted signal, we plot in Figs.~\ref{fig:anhar}b-c the second-order correlation function for the vibration and the filtered anti-Stokes field as functions of $\frac{n_{\rm coh}}{n_{\rm th}}$ and $\frac{\chi \omega_{\rm b}}{\gamma}$. Overall, these color maps show that antibunching of the filtered anti-Stokes field is achieved by having strong enough MIR drive and large enough anharmonicity (region {C} in both plots). For weak anharmonicity (regions {A} and {B}), an increase in MIR drive causes a transition from a thermal to a coherent state for both the vibration and the anti-Stokes field, as already discussed in \autoref{fig:harmo}b. Finally, in region {D}, where the anharmonicity is large but the MIR drive is weak, the vibrational mode remains in thermal equilibrium but the filtered anti-Stokes photons may become antibunched as a result of rejecting all photons originating from Raman transitions beyond the ground to first excited vibrational state. The benefit of the MIR drive in our scheme is evidenced in Fig.~\ref{fig:anhar}c by the higher degree of antibunching in region {C} with less stringent requirements on the magnitude of anharmonicity, compared to region D. 

\begin{figure}[h!]
\centering
\includegraphics[width=\columnwidth]{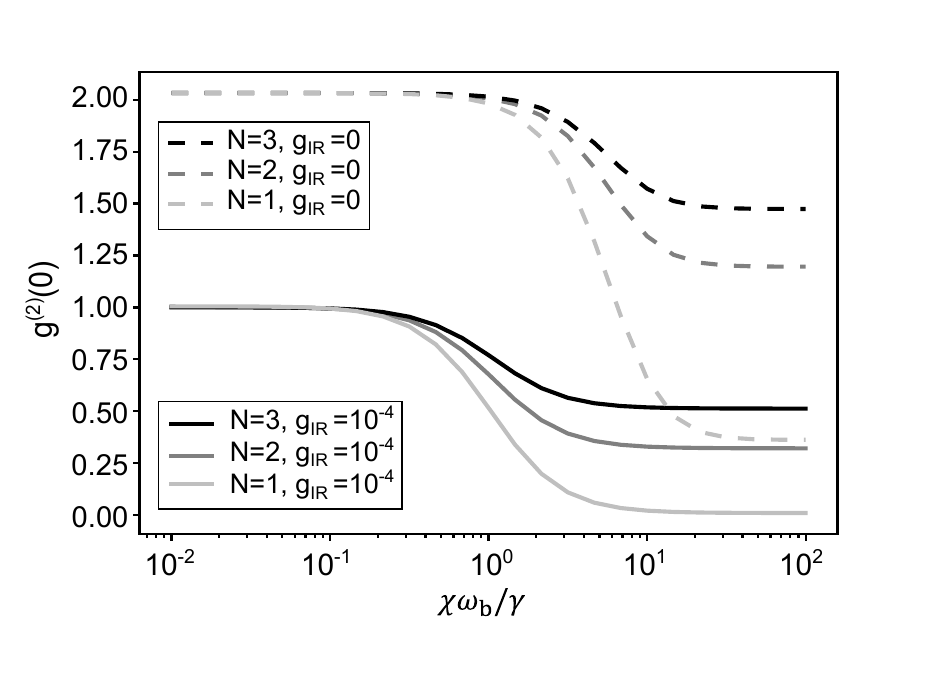}
\caption{Second-order correlation function of anti-Stokes plotted versus anharmonicity for different number of molecules.}
\label{fig:many}
\end{figure}

\paragraph{Many Molecules}

Achieving single molecule Raman spectroscopy remains a nontrivial experimental task despite multiple demonstrations since its first observation in 1997~\cite{kneipp1997single}. We therefore investigate whether antibunching persists in the presence of several molecules that contribute to VSFG\@. In the limit of a large number of molecules, collective vibrational excitations are expected to become perfectly harmonic, which is a general result for an ensemble of two-level systems~\cite{gross2012spin}. But in our system the increase in the number of molecules leads to an increase in the collective resonant (MIR) and optomechanical (VIS/NIR) coupling strengths, potentially resulting in the excitation of higher excited vibrational states thereby recovering the anharmonicity. To clarify the expected trend, we perform complete numerical evaluation of the model with $N=1,2$ and $3$ molecules. For simplicity, we consider that all molecules are identical and have the same coupling rates to the common nanocavity modes:
\begin{equation}
\label{eq: manymolecule}
\begin{split}
H_N =\ & (\omega_{\rm c}- \omega_{\rm p})\delta a_{\rm c}^{\dagger} \delta a_{\rm c} + \sum_{i=1}^{N}\omega_{\rm b} b^{\dagger}_i b_i - \frac{\chi}{2} \omega_{\rm b} b^{\dagger}_i b^{\dagger}_i b_i b_i\\
& + \sum_{i=1}^{N} g_{\rm c} ( \delta a_{\rm c}^{\dagger} + \delta a_{\rm c} )(b^{\dagger}_i + b_i )\\
& + \sum_{i=1}^{N} g_{\rm IR} ( b_i e^{i \omega_{\rm d} t}+ b^{\dagger}_i e^{-i \omega_{\rm d} t})
\end{split}
\end{equation}
The second-order correlation $g^{(2)}(0)$ for the filtered anti-Stokes field obtained from the solution to the corresponding Lindblad master equation is plotted in \autoref{fig:many} for $N=1,2,3$, with (solid lines) and without (dashed lines) an MIR drive. The results confirm the intuition from the large $N$ limit: the degree of achievable antibunching quickly decreases as the number of molecules participating in Raman scattering and VSFG increases. Interestingly, we find that this decrease cannot be compensated by arbitrarily increasing the anharmonicity of each molecule as characterized by the parameter $\chi$. This can be attributed to the fact that for large enough anharmonicity, each vibration acts as a two-level system, and the collective behavior of the ensemble is dominated by the harmonic limit mentioned above.

\section{Conclusion}

We have proposed a new scheme to produce antibunched photons by performing vibrational sum-frequency generation on a single molecule. Our method leverages the anharmonicity naturally present in the potential of vibrational modes that are localized on few molecular bonds to realize a form of conventional photon blockade without the need for strong light-matter coupling. Our calculations show that under sufficiently strong MIR drive, characterized by the ratio of coherent to thermal vibrational population $\frac{n_\text{coh}}{n_\text{th}}\geq 10$, almost complete antibunching is achieved whenever the anharmonicity satisfies $\chi\omega_{\rm b}\geq\gamma$, i.e., the energy difference between the first and second vibrational transitions is larger than the level linewidth. 

Our proposal is feasible with current experimental capabilities. The main requirement is to be able to measure Raman scattering from a single molecule, which can be achieved in various plasmonic nanoantennas and scanning tip systems~\cite{qiu2022singlemolreview}. Efficient coupling of MIR light into the system can be achieved using dual-resonant plasmonic antennas~\cite{chen2021continuous,xomalis2021} or the broadband nanofocusing capabilities of metallic tips~\cite{huth2012nanoFTIR,pechenezhskiy2013}. The main foreseen difficulty will be to spectrally and temporally isolate the antibunching dip in the anti-Stokes photon flux: First, one should expect a braodband non-resonant nonlinear sum frequency signal coming from the metallic surface and the molecule's electronic response~\cite{himmelhaus2000}, which should ideally be kept much smaller than the vibrational contribution to SFG. Second, the time resolution needed to resolve antibunching is on the order of one picosecond, the typical relaxation time for molecular vibrations on a substrate~\cite{jakob2024}; since the fastest single photon detectors to date feature jitters above 10~ps, the experiment should therefore be conducted under pulsed excitation. Overall, none of these challenges seem insurmountable. The richness and versatility of the proposed scheme may also open new perspectives for the generation of entangled photons and quantum metrology applications based on the detection of photon correlations.

\section*{Acknowledgements}
C.G. and F.M.K. acknowledge Hanyuan Hu for initial code development related to this study. This work has received funding from the European Union's Horizon 2020 research and innovation program under Grant Agreement No. 820196 (ERC CoG QTONE) and from the Swiss National Science Foundation (project numbers 214993 and 198898).

\bibliography{References,VibrationalSumFrequency}

\newpage
\appendix

\section{Beam splitter interaction}
\label{subsec: beamsplitter interaction}
\begin{figure}[H]
\centering
\includegraphics[width=0.5\columnwidth]{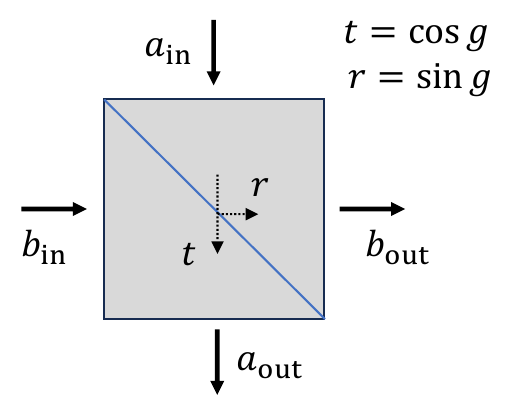}
\caption{Beam splitter with transmission and reflection coefficients of $t$ and $r$ and two input arms and two output arms.}
\label{fig: BS}
\end{figure}

Assuming that the optomechanical coupling is small compared to the infrared coupling in \autoref{eq: finalHamilton}, $g_{\rm IR} \gg g_c$, one can neglect quantum backaction of the cavity mode on the vibration, whose state is then mostly governed by the coupling to the MIR field and to the environment.
To estimate the state of optical mode at the anti-Stokes sum-frequency sideband of the VIS/NIR laser, we need to treat the optomechanical coupling term, $g_{\rm c} ( \delta a_{\rm c}^{\dagger} + \delta a_{\rm c} )(b^{\dagger} + b )$, which reduces to  $g_{\rm c} ( \delta a_{\rm c}^{\dagger}b+ \delta a_{\rm c} b^{\dagger} )$ in the rotating wave approximation. It is a bilinear interaction between the two modes and, after changing the notation $\delta a_{\rm c}\longrightarrow a$, the associated unitary operator takes the form
$$U_{\rm BS}= e^{g(e^{i \theta} a^{\dagger} b - e^{-i \theta} a b^{\dagger})}$$
where $g= \norm{g_c}$ and $\theta =\arg{(g_c)}$. 
It is a general beam splitter interaction as depicted in \autoref{fig: BS} with the relation between input and output operators given by
\begin{equation}
\begin{split}
    a_{\rm out} &= U_{\rm BS}^{\dagger}\, a_{\rm in} \, U_{\rm BS} = \cos(g) a_{\rm in} + \sin(g) e^{i \theta} b_{\rm in} \\
    b_{\rm out} &= U_{\rm BS}^{\dagger}\, b_{\rm in}\, U_{\rm BS} = \cos(g) b_{\rm in} - \sin(g) e^{-i \theta} a_{\rm in}
\end{split}
\label{eq: BS}
\end{equation}
where $g$ characterizes the beam splitter transmission ratio through $t=\cos{g}$ and $r=\sin{g}$, see \autoref{fig: BS}.

In our context, the input modes of the beam splitter are populated by vacuum fluctuations for the anti-Stokes field $a_{\rm in}$ and by the thermal and MIR-driven vibrational state for $b_{\rm in}$; the output mode $a_{\rm out}$ corresponds to the emitted anti-Stokes field on which the $g^{(2)}$ measurement is performed, while $b_{\rm out}$ reflects the state of the vibration modified by the anti-Stokes part of the optomechanical interaction, but it is a weak perturbation since we neglect quantum backaction. Finally, a known result is that, when one of the input mode is in a vacuum state, the $g^{(2)}$'s of both outgoing fields from a beam splitter are equal to that of the non-vacuum input port, in particular here $g^{(2)}_{a_{\rm out}}=g^{(2)}_{b_{\rm in}}$ (see e.g. Sec. IV.D in Ref.~\citenum{campos1989beamsplitter}). In other word, the vibrational mode excitation number statistics is mapped onto the anti-Stokes field generated through the optomechanical (Raman) interaction. The (usually dominant) vacuum noise entering in the process has no impact on photon number detection, as long as dark counts can be neglected~\cite{sekatski2012}.


\section{Frequency-filtered correlation function}
\label{subsec: freqFilteredMethod}
The fluctuation of the optical mode, $\delta a_c$, encompasses both Stokes and anti-Stokes components. To exclusively examine the second-order correlation function of a specific component, one can define a pair of two-level systems (filters or sensors) at the component of interest's frequency and weakly couple them to $\delta a_c$ by a beam-splitter interaction~\cite{delvalle_2012}. The second-order correlation function is then given by the correlation between the sensors' populations.

As an illustration for the anti-Stokes component, the following Hamiltonian is added to the system's Hamiltonian in \autoref{eq: finalHamilton}:
\begin{equation}
    \begin{split}
        H_{\rm Sens} &= \omega_{\rm aS}\zeta^{\dagger}_1 \zeta_1 + \omega_{\rm aS}\zeta^{\dagger}_2 \zeta_2, \\
        H_{\rm Coupl} &= \epsilon_{\rm }(\delta a_c \zeta^{\dagger}_1+\delta a^{\dagger}_c\zeta_1) + \epsilon(\delta a_c \zeta^{\dagger}_2+\delta a^{\dagger}_c\zeta_2),
    \end{split}
\end{equation}
where $\zeta_1$ and $\zeta_2$ are bosonic annihilation operators of the two level systems, each having a decay rate of $\Gamma$ that defines the filter linewidth in our model. To maintain the system's solution unperturbed, $\epsilon$ must be sufficiently small, satisfying the condition $\epsilon \ll \sqrt{\frac{\gamma \Gamma}{2}}$. Subsequently, $g^{(2)}_{\rm aS}(0)$ is expressed as follows:
\begin{equation}
    g^{(2)}_{\rm aS}(0)= \frac{\langle \zeta^{\dagger}_1 \zeta_1 \zeta^{\dagger}_2 \zeta_2 \rangle}{\langle \zeta^{\dagger}_1 \zeta_1 \rangle \langle \zeta^{\dagger}_2 \zeta_2 \rangle}
\end{equation}
For further details, please refer to~\cite{delvalle_2012}.  

\section{Impact of filter linewidth on measured $g^{(2)}(0)$}

\begin{figure}
    \centering
    \includegraphics[width=\columnwidth]{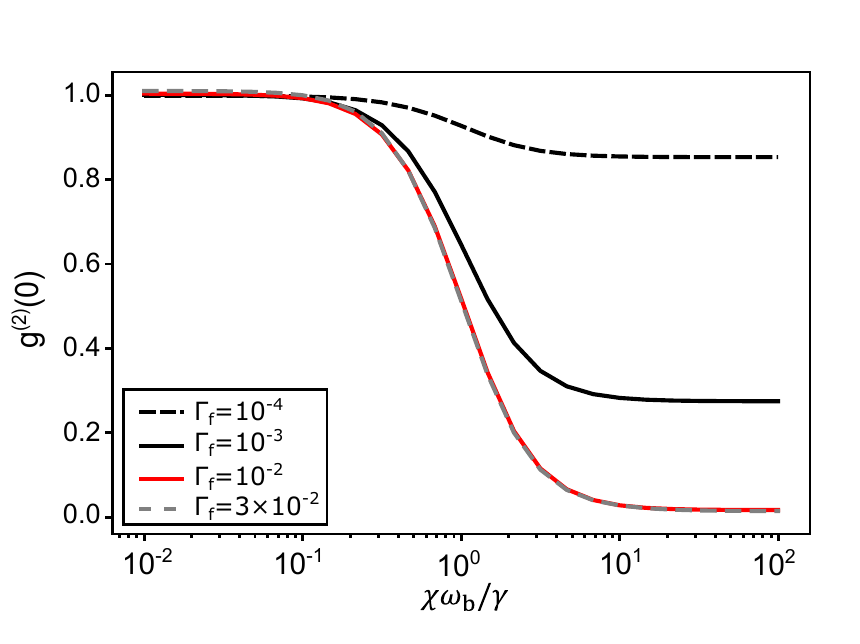}
    \caption{$g_{\rm aS}^{(2)}(0)$ variation with anharmonicity in the high infrared drive amplitude regime, considering different filter linewidths.}
    \label{fig:g2_vs_chi_diff_Gamma_highIR}
\end{figure}

As investigated in Ref.~\citenum{carreno2022loss}, too narrow spectral filtering may alter the intrinsic photon statistics of a single photon source into that of a thermal state. In our study, under strong MIR-drive, the intensity of the anti-Stokes field is dominated by VSFG over spontaneous Raman scattering, there is no need to aggressively filter the signal; the role of filtering is here to suppress the contribution to $g^{(2)}$ from spontaneous Stokes scattering and VDFG (difference frequency) on the other side of the VIS/NIR laser. 
We show in \autoref{fig:g2_vs_chi_diff_Gamma_highIR} that, as found in Ref.~\citenum{carreno2022loss}, a decrease in filter linewidth destroys antibunching in the filtered field. That's why we chose $\Gamma = 10 \gamma = 10^{-2}$ for all figures in the main text under MIR drive. 

\begin{figure}
    \centering
    \includegraphics[width=\columnwidth]{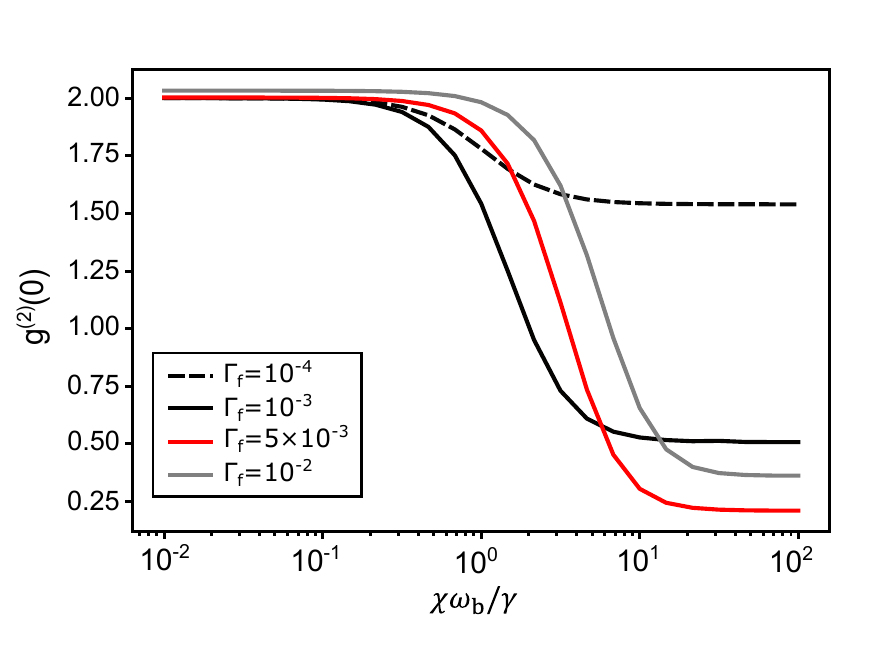}
    \caption{$g_{\rm aS}^{(2)}(0)$ variation with anharmonicity in the low infrared drive amplitude regime, considering different filter linewidths.}
    \label{fig:g2_vs_chi_diff_Gamma_lowIR}
\end{figure}

In the absence of MIR drive, photon antibunching is only obtained by rejecting anti-Stokes photons coming from the second (and higher) vibrational transition, which is detuned by $\chi\omega_{\rm b}$ from the first one, imposing an upper limit on filter linewidth $\Gamma\leq \chi\omega_{\rm b}$. Besides, the constraint $\chi\omega_{\rm b}\geq\gamma$ remains, as for any conventional photon blockade scheme, leading to $\gamma,\Gamma\leq \chi\omega_{\rm b}$. Finally, as reminded above, filtering below the natural linewidth of the anti-Stokes photons $\gamma$ degrades antibunching by `thermalizing' the photon statistics, imposing $\Gamma\geq \gamma$. Altogether we obtain the stringent constraints $\gamma\leq \Gamma\leq \chi\omega_{\rm b}$, which leaves little room for the choice of $\Gamma$ since in practice the anharmonicity is not much greater than $\gamma$. This parameter space is explored in \autoref{fig:g2_vs_chi_diff_Gamma_lowIR}. Overall, for a fixed filter linewidth, the transition to an antibunched state occurs when $\chi \approx \frac{\Gamma}{2}$. But as the filter linewidth is decreased below $\gamma$, $g^{(2)}(0)$ approaches 2 even where antibunching is expected, consistent with the findings of Ref.~\citenum{carreno2022loss}. 

\section{Approximations}
To derive \autoref{eq:linearizedHamilton}, several approximations are employed.

First, the rotating wave approximation is applied to the MIR coupling term. Second, both the MIR and VIS/NIR modes are linearized since their coupling rates to the vibration are both assumed to be small compared to all other energies in the Hamiltonian. 
Consequently, $a_c$ is expressed as $\alpha_c + \delta a_c$, where $\alpha_c$ is the solution in the absence of the optomechanical coupling term, representing a coherent state, and $\delta a_c$ represents the fluctuations induced by the optomechanical coupling term. Additionally, $a_{\text{IR}}$ is replaced by $\alpha_{\text{IR}}$, with $\alpha_{\text{IR}}$ representing the coherent state in the absence of the MIR coupling term. The fluctuations in this case are neglected for simplicity, as the primary interest lies in the VSFG field properties contained in $\delta a_c$.

Next, since the simplified Hamiltonian, \autoref{eq: finalHamilton}, is periodically time-dependent, a Floquet solution of the following form has been considered for the Lindbladian master equation:
\begin{equation}
    \label{eq:DMexpansion}
    \rho(t) = \sum_{\rm n= - n_b}^{ n_b} \rho_{\rm n} e^{i n \omega_{\rm d} t}
\end{equation}
Where in principle the bounds of the sum must go to infinity. Nevertheless, we truncate at $n_b=4$ and it is verified that considering more terms does not change the results.

\end{document}